\documentclass[10pt]{iopart}
\usepackage{epsfig,amssymb}
\begin{document}

\title{Kaon and Antikaon Production in Heavy Ion Collisions at 1.5 AGeV}

\author{Andreas F{\"o}rster$^{\star}$ for the KaoS Collaboration\footnote[1]{H.Oeschler, 
C.Sturm, F.Uhlig (TU Darmstadt), P.Koczo{\'n}, E.Schwab, P.Senger (GSI Darmstadt), 
Y.Shin, H.Str{\"o}bele (University of Frankfurt), I.B{\"o}ttcher, B.Kohlmeyer, M.Menzel, 
F.P{\"u}hlhofer (University of Marburg), W.Walu{\'s} (Jagellonian University Cracow), 
E.Grosse, L.Naumann, W.Scheinast, A.Wagner (FZ Rossendorf)} }

\address{$^{\star}$Institut f{\"u}r Kernphysik, 
           Technische Universit{\"at} Darmstadt,\\
            D-64289 Darmstadt, Germany}

\ead{a.foerster@gsi.de}

\begin{abstract}

At the Kaon Spectrometer KaoS at SIS, GSI the production of kaons and antikaons in heavy 
ion reactions at a beam energy of $1.5$~AGeV 
has been measured for the collision systems Ni~+~Ni and Au~+~Au. 
The $K^{-}/K^{+}$ ratio is found to be constant for both systems and as a function of impact parameter
but the slopes of $K^{+}$ and $K^{-}$ spectra differ for all impact parameters.
Furthermore the respective polar angle distributions will be presented as a function of centrality.

\end{abstract}

\section{Introduction}

Heavy ion collisions provide the unique possibility to study strange mesons 
in baryonic matter at densities well above saturation density. 
Of special interest is their production and propagation at and below the 
NN-threshold \cite{ka_kaos,ka_fopi}.
At higher beam energies (AGS, SPS, RHIC) the abundances of produced particles can be well described 
within the framework of a statistical model \cite{pbm}. The same holds 
for SIS energies if strangeness conservation is treated canonically \cite{cley}.
In the following 
$K^{+}$ and $K^{-}$ spectra measured recently in Au~+~Au and Ni~+~Ni collisions at $1.5$~AGeV
with the Kaon Spectrometer \cite{se93} at SIS/GSI
will be presented as a function of centrality as well as the respective polar angle distributions. 
A general overview of KaoS results
is given in the talk of C.Sturm.

\section{Polar Angle Distribution}

\begin{figure}
\noindent
\begin{minipage}[t]{5.3cm}
  \centering \epsfig{figure=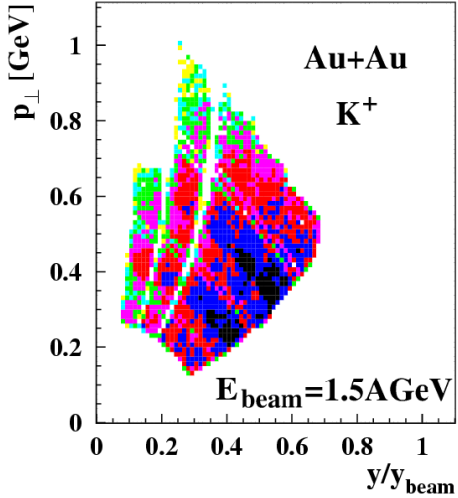,width=7cm,clip=}
\end{minipage} \hfill
\begin{minipage}[b]{7cm}
   \caption{Transverse momentum as a function of rapidity for $K^{+}$. 
             The different bands correspond 
             to different spectrometer angles in the laboratory.}
    \label{fig_pty}
\end{minipage} 
\end{figure}

By pivoting the Kaon Spectrometer around the target point 
a wide range in polar angle within the center of momentum frame is covered. 
Figure \ref{fig_pty} shows the range in transverse momentum and rapidity for $K^{+}$  
in Au~+~Au collisions at 1.5 AGeV beam energy covered by five angular settings in the laboratory.

\begin{figure}
\noindent
\begin{minipage}[t]{5.3cm}
  \mbox{\epsfig{figure=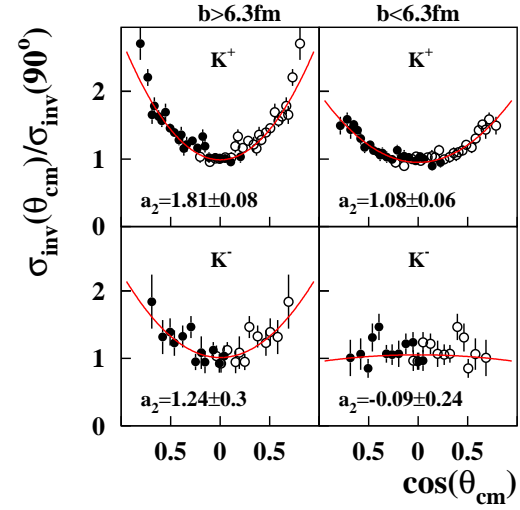,width=7cm,clip=}}
\end{minipage} \hfill
\begin{minipage}[b]{6.5cm}
  \caption{Preliminary polar angle distributions for $K^{+}$ and $K^{-}$ in Au~+~Au 
           collisions at $E_{beam}=1.5$~AGeV. The left panel shows data for impact 
           parameters $b>6.3$~fm, the right one for $b<6.3$~fm. Fits and the parameter $a_{2}$ 
           are according to equation \ref{eq_2}.}
  \label{fig_angdis}
\end{minipage} 
\end{figure}

Figure \ref{fig_angdis} depicts the polar angle distribution 
$\sigma_{inv} \left( \theta_{cm} \right) \, /  \, \sigma_{inv} \left( 90^{\circ} \right)$
for $K^{+}$ and $K^{-}$ for two centrality classes.
The lines represent fits 
assuming a quadratic dependence on $\cos(\theta_{cm})$:

\begin{equation} \label{eq_2}
\frac{d \sigma}{ d \left( \cos \theta_{cm} \right)} \quad \sim \quad 
   1 \, + \, a_2 \cdot \cos^{2}\left( \theta_{cm} \right)   \quad .
\end{equation}

This anisotropy is more pronounced for $K^{+}$ than for 
$K^{-}$ and it decreases with increasing centrality. Near central $K^{-}$ data are isotropic.
Similar trends have been observed in Ni~+~Ni collisions at a beam energy of 1.93~AGeV \cite{me00}.

\section{Centrality Dependence}

\begin{figure}
\noindent
\begin{minipage}[t]{5.3cm}
  \mbox{\epsfig{figure=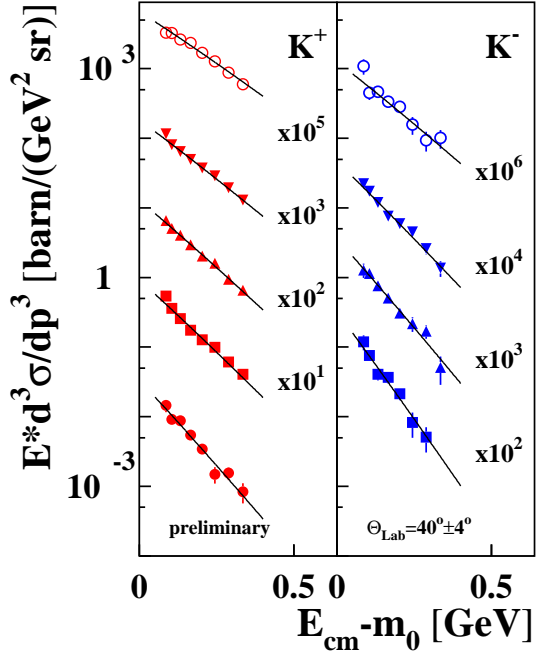,width=7cm,clip=}}
\end{minipage} \hfill
\begin{minipage}[b]{6.5cm}
  \caption{Invariant cross sections for $K^{+}$ and $K^{-}$ in Au~+~Au collisions 
            at $E_{beam}=1.5$~AGeV for 5 centrality classes. The open circles depict 
            the most central data with decreasing centrality from the top to the bottom 
            of the figure.}
  \label{fig_spec_tbin}
\end{minipage} 
\end{figure}

To investigate the centrality dependence of the kaon and antikaon production we have divided 
the data measured close to midrapidity $(\theta_{lab}=40^{\circ})$ into five centrality 
classes. Figure \ref{fig_spec_tbin} shows the corresponding energy spectra 
for Au~+~Au collisions at $E_{beam}=1.5$~AGeV. The lines are Boltzmann fits.
The resulting inverse slope parameters for the $K^{+}$ spectra are about $20$~MeV higher 
than those for the $K^{-}$ spectra for all centrality classes. 

\begin{figure}[t]
 \begin{center}
  \mbox{\epsfig{file=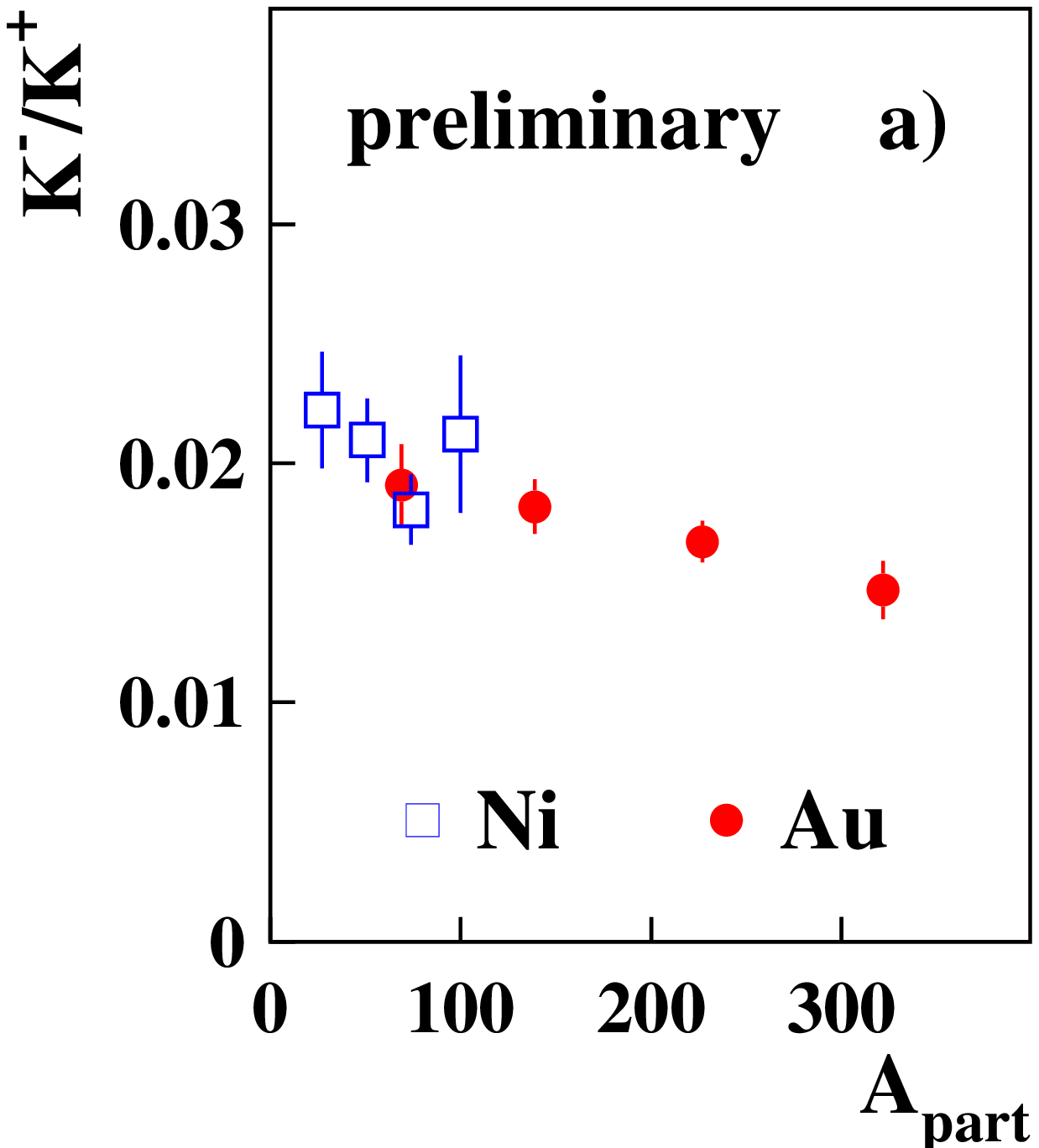,width=4.2cm,clip=}}
  \mbox{\epsfig{file=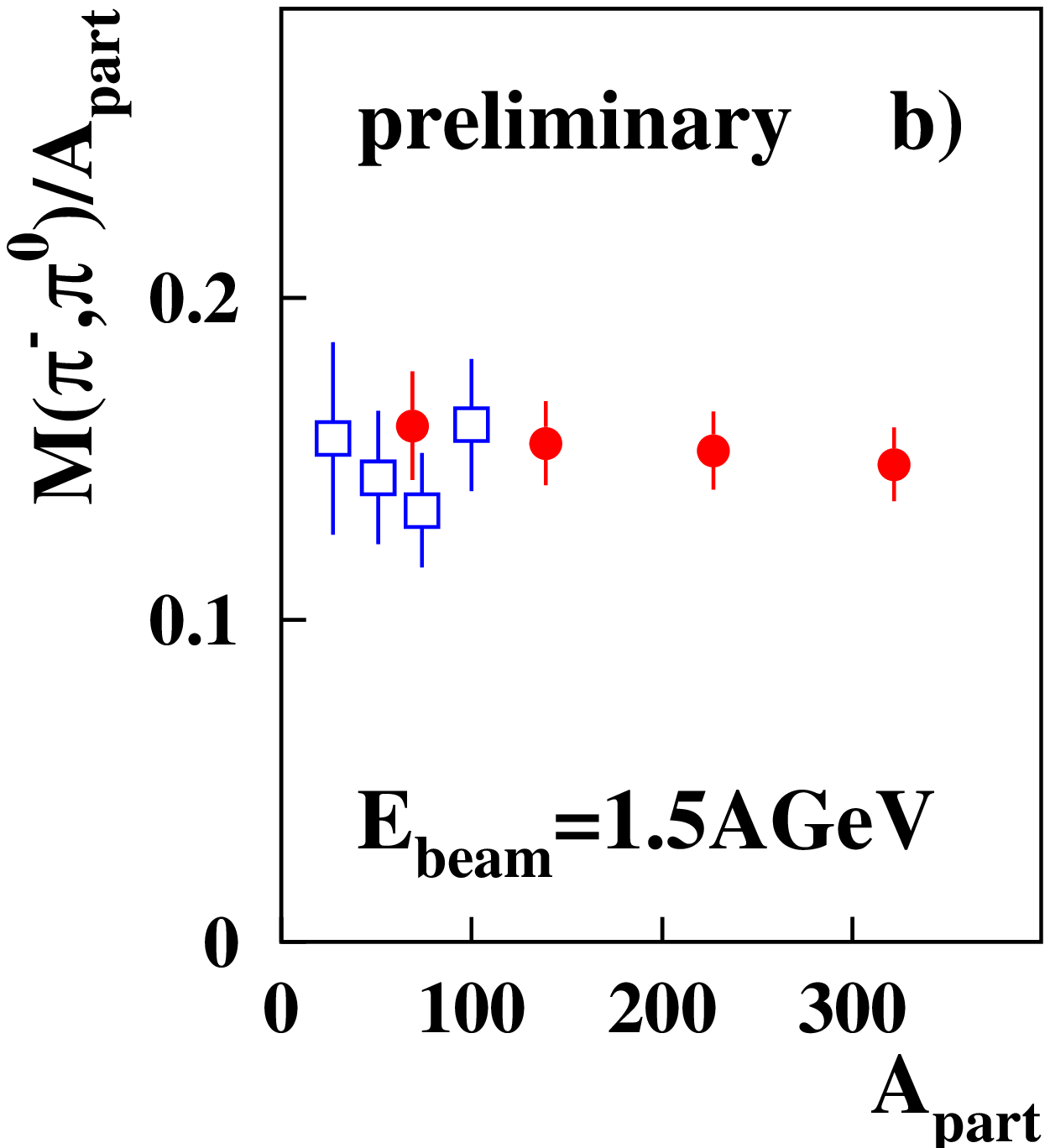,width=4.2cm,clip=}}
  \mbox{\epsfig{file=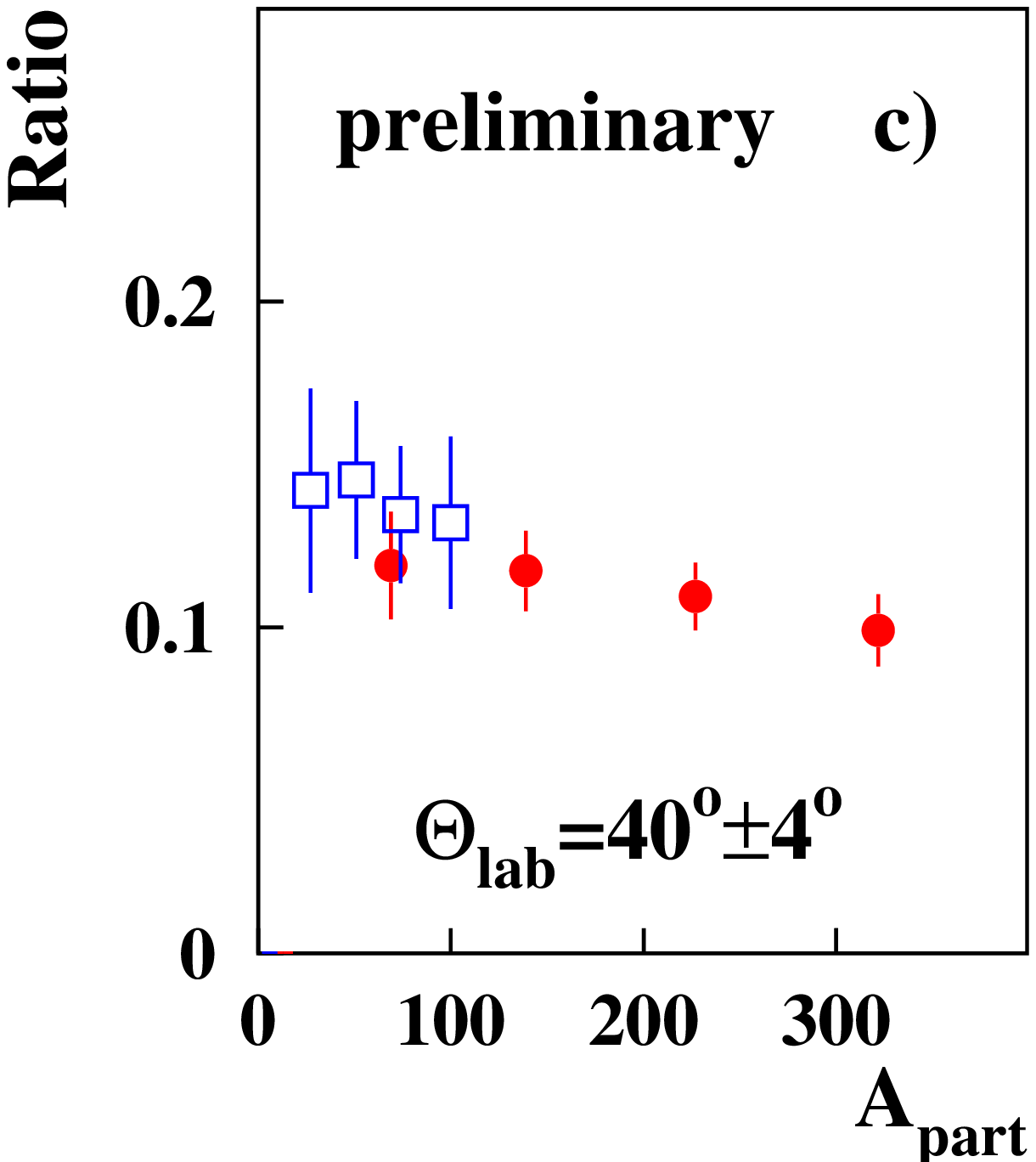,width=4.2cm,clip=}}
  \caption{Part a) shows the $K^{-}/K^{+}$ ratio, part b) the multiplicity 
           of $\pi^{-}$ and $\pi^{0}$ per $A_{part}$ and part c) the double ratio 
           derived by dividing a) by b) as a function of $A_{part}$.}
  \label{fig_triplepic}
 \end{center}
\end{figure}

The dependence of the particle multiplicity on the number of participating nucleons $A_{part}$ 
as determined from the measured centrality by a geometrical model is found to be the same for 
kaons and antikaons. This is also observed in the Ni~+~Ni case. Even the absolute 
values of the multiplicity per $A_{part}$ at a given $A_{part}$ are the same 
for Au and Ni. The $K^{+}$ and $K^{-}$ production seems to depend only 
on the number of participating nucleons. 
As a result the $K^{-}/K^{+}$ ratio is constant as a function of $A_{part}$ 
independent of the collision system (figure \ref{fig_triplepic} a) indicating a link 
between the production mechanisms of both particles. 

As suggested in \cite{ko84} and supported by recent transport calculations \cite{ha01} 
the dominant production channel for $K^{-}$ is the strangeness exchange reaction
 
\begin{equation} \label{eq_3}
\pi^{(0,-)} \, + \, Y \, \leftrightharpoons \, K^{-} \, + \, N  \quad .   
\end{equation}

If one naively assumes this channel to be in chemical equilibrium the 
law of mass action could be applied \cite{oe00}:  

\begin{equation} \label{eq_4}
\frac{[\pi^{(0,-)}] \cdot [Y]}
{[K^{-}] \cdot [N]} \, = \, \kappa \, = \, const.
\end{equation}

At this beam energy most hyperons $Y$ are produced associately with a 
kaon.
Since the number of hyperons is changed only marginally by the production of 
$K^{-}$ their rate can be substituted by the one of the $K^{+}$.
Using additionally $A_{part}$ for $[N]$ yields 

\begin{equation} \label{eq_5}
\frac{[K^{-}] / [K^{+}]}{[\pi^{(0,-)}] / A_{part}} \, 
  \sim  \, \frac{1}{\kappa} \, = \, const. 
\end{equation}

Figure \ref{fig_triplepic} b) shows the multiplicity of $\pi^{-}$ and $\pi^{0}$ per $A_{part}$ 
(with $M(\pi^{0}) = 0.5 \cdot [M(\pi^{+})+M(\pi^{-})]$) as a function of $A_{part}$.
Figure \ref{fig_triplepic} c) shows the double ratio of equation \ref{eq_5} obtained by
dividing the $K^{-}/K^{+}$ ratio from figure \ref{fig_triplepic} a) by the pion multiplicity per 
$A_{part}$ from figure \ref{fig_triplepic} b).
The result seems to be a constant value independent of the number of participating 
nucleons and the collision system.

\section{Conclusions}

The constancy of the double ratio $(K^{-}/K^{+}) \, / \, (M(\pi^{-},\pi^{0})/A_{part})$ 
nicely agrees with the assumption of the strangeness exchange reaction being the dominant 
production process for $K^{-}$ at SIS energies. 
While this supports the idea of chemical equilibration in this specific channel 
the differences in the spectral slopes of $K^{+}$ and $K^{-}$ are in contradiction.

\end{document}